\documentclass[a4paper] {article}

\usepackage{amsmath}
\usepackage{color}
\newcommand{\ra}{\rightarrow}

\newcommand{\R}{\mathcal{R}}

\newcommand{\N}{\mathcal{N}}

\usepackage{latexsym}
\usepackage{graphicx}
\usepackage{color}
\usepackage{float}

\begin{document}
\begin{center}
\textbf{\Large DISTRIBUTED REINFORCEMENT\\ \ \\ LEARNING VIA GOSSIP}
\end{center}

\vspace{1in}

\begin{center}
ADWAITVEDANT S.\ MATHKAR AND VIVEK S.\ BORKAR\footnote{Research supported in part by a J.\ C.\ Bose Fellowship and a grant `Distributed Computation for Optimization over Large Networks and High Dimensional Data Analysis' from the Dept.\ of Science and Technology, Govt.\ of India.}\\
Department of Electrical Engineering, \\
Indian Institute of Technlogy, \\
Powai, Mumbai 400076, India.\\
(mathkar.adwaitvedant, borkar.vs@gmail.com)
\end{center}

\vspace{.5in}

\noindent \textbf{\large Abstract:} We consider the classical TD(0) algorithm implemented on a network of agents wherein the agents also incorporate the updates received from neighboring agents using a gossip-like mechanism. The combined scheme is shown to converge  for both discounted and average cost problems.\\\


\section{Introduction}

Reinforcement learning with function approximation has been a popular framework for approximate policy evaluation and dynamic programming for Markov decision processes (Bertsekas 2012; Gosavi 2003; Lewis and Liu 2013; Powell 2007; Szepesvari 2010). In view of the growing interest in control across communication networks, there has been a growing need to consider distributed or multi-agent versions of these schemes. While there has been some early work in this direction, analysis of  provably convergent schemes is lacking. (See, e.g., Lauer and Riedmiller (2000), Littman and Boyan (1993), Pendrith (2000), Weiss (1995), also Busoniu et al (2008) and Panait and Luke (2005) for surveys. The work closest to ours in spirit is Macua et al (2012).) The present article aims at filling up this lacuna. Specifically, we consider a distributed version of the celebrated TD(0) algorithm implemented across a network of processors or `agents', who communicate with each other and incorporate, in addition to their own measurements, the estimates of their neighbors. For the latter aspect, we borrow a simple averaging scheme from gossip algorithms (Shah, 2008). We prove the convergence of this scheme. It may be noted that we \textit{do not} prove consensus, in fact consensus is an unreasonable expectation here. This is so because each agent potentially has a different set of basis functions, even of a different cardinality.  We do, however, justify the proposed scheme in terms of a certain performance measure.\\

Next section describes our convergence results for the infinite horizon discounted cost problem. Section 3 extends them to the average cost problem. Section 4  comments upon the results.\\

\section{Discounted cost}

Let $\{X_n\}$ denote  an irreducible  Markov chain on a finite state space $S := \{1, 2, \cdots, m\}$ with transition matrix $P := [[p(i,j)]]_{i, j \in S}$, and an associated  `running' cost function $c: S \times S \times \R$. Thus $c(i,j)$ denotes the cost associated with the transition from $i$ to $j$. (While we have a controlled Markov chain in mind, we are interested in estimating the cost for a fixed policy, so we do not render explicit the policy dependence of $P, c$ for sake of notational ease.) Consider the problem of estimating the infinite horizon cost
\begin{displaymath}
J(x) := E_x\left[\sum_{t = 0}^{\infty}\alpha^mc(X_t, X_{t + 1})\right],
\end{displaymath}
 $\alpha \in (0, 1)$ being the discount factor. The original $TD(0)$ algorithm  for approximate evaluation of $J$ begins with the a priori approximation $J(\cdot) \approx \phi(\cdot) r = \sum_{i=1}^Kr_i\phi_i(\cdot)$. Here ${\phi}(\cdot) = [{\phi}_{1}(\cdot) : {\phi}_{2}(\cdot) : \cdots : {\phi}_{K}(\cdot)]^{T}$ with ${\phi}_{i} :=$  the $ith$ feature vector (these are kept fixed), and $r := [r_1, \cdots, r_K]^T$ are the weights that are to be learnt. The actual algorithm for doing so is as follows (Tsitsiklis and Van Roy, 1997):
\begin{equation}
r_{t+1} = r_t + {\gamma}_t {\phi}(X_t) [c(X_t,X_{t+1}) + {\alpha}{\phi}(X_{t+1})^{T}r_{t} -{\phi}(X_t)^{T}r_{t}],
\end{equation}
where the step-sizes $\gamma_t > 0$ satisfy $\sum_t\gamma_t = \infty, \sum_t\gamma_t^2 < \infty$. A convergence proof and error estimates relative to the exact $J$ may be found in Tsitsiklis and Van Roy (1997). We sketch an alternative convergence proof of independent interest, using the `o.d.e.' (for \textit{ordinary differential equation}) approach of Derevitskii and Fradkov (1974) and Llung (1977). For simplicity, we rely on the exposition of Borkar (2008). Let $\eta$ denote the unique stationary probability vector for the chain and $D$ the diagonal matrix whose $i$th diagonal entry is $\eta(i)$. By Corollary 8, p.\ 74, Borkar (2008), the `limiting o.d.e.' for the above iteration is
\begin{equation}
\dot{r}(t) = {\phi}^{T}{D} \bar{c} + {\alpha}{\phi}^{T}{D}{P}{\phi}r(t) - {\phi}^{T}{D}{\phi}r(t) := h(r(t)) \label{ode0}
\end{equation}
for $h(x) := {\phi}^{T}{D} \bar{c} + {\alpha}{\phi}^{T}{D}{P}{\phi}x - {\phi}^{T}{D}{\phi}x, \ \bar{c} := \sum_jp(i, j)c(i, j).$ Then
\begin{displaymath}
h_{\infty}(x) := \lim_{a\uparrow\infty}\frac{h(ax)}{a} = {\alpha}{\phi}^{T}{D}{P}{\phi}x - {\phi}^{T}{D}{\phi}x.
\end{displaymath}
It is  easy to see that $\frac{1}{a}h(ax) \rightarrow h_{\infty}(x)$ uniformly on $R^{K}$.\\

\textbf{Theorem 1:} Under the above assumptions, $r_t \ra \hat{r}$ a.s., where $\hat{r}$ is the unique solution to $h(\hat{r}) = 0$.\\

\textbf{Proof:} The `scaled o.d.e.' $\dot{r}(t) = h_{\infty}(r(t))$ is a linear system with the origin as its globally asymptotically stable equilibrium, in fact $V(r(t)) = \|r(t)\|^2$ is a Liapunov function, as seen from Lemma 9 of Tsitsiklis and Van Roy (1997) with $r^*$ therein replaced by the zero vector. By Theorem 9, p.\ 75, Borkar (2008), $\sup_t\|r_t\| < \infty$ a.s. In turn, (\ref{ode0}) has $\hat{r}$ as its globally asymptotically stable equilibrium, again $V(r(t)) = \|r(t) - \hat{r}\|^2$ is a Liapunov function, as seen from Lemma 9 of Tsitsiklis and Van Roy (1997). The claim follows by Theorem 7 -- Corollary 8, p.\ 74, Borkar (2008). \hfill $\Box$\\

We now describe the distributed version of this scheme. Consider $n$ agents sitting on the nodes of a connected graph, each with a different set of feature vectors. We denote by $\N(i)$ the set of neighbors of $i$. Let the feature vectors of the $ith$ agent be denoted by
 ${\phi}^{i}_{1}, {\phi}^{i}_{2},......{\phi}^{i}_{n_i} $, with  $\Phi^i := [{\phi}^{i}_{1} : {\phi}^{i}_{2} : ...... : {\phi}^{i}_{n_i}]^T$.
Let $q(i,j)$ denote the probability by which  $ith$ agent polls agent $j \in \N(i)$. The $ith$ agent runs the following $n_i$-dimensional iteration:
\begin{equation}
r^{i}_{t+1} = r^{i}_t + {\gamma}_t {\phi}^{i}(X_t) [c(X_t,X_{t+1}) + {\alpha}{\phi}^{Y^{i}_{t+1}}(X_{t+1})^{T}r^{Y^{i}_{t+1}}_{t} -{\phi}^{i}(X_t)^{T}r^{i}_{t}]. \label{iter-i}
\end{equation}
Here $Y^{i}_{t}$ is a $[1,2...n]$ valued random variable taking value j with probability $q(i,j)$ . We further assume it is independent of $\{X_s, Y^j_s, j \neq i, s \leq t; Y_s^i, s < t\}$.

We make the following key assumptions:

\begin{itemize}

\item $\textbf{(A1)}$  ${ \phi}^{i}_{1}, {\phi}^{i}_{2},......{\phi}^{i}_{n_i}$ are linearly independent for all $i$.

\item $\textbf{(A2)}$
The Markov chain  $\{X_t\}$ is irreducible and aperiodic.

\item $\textbf{(A3)}$
The stochastic matrix $Q := [[q(i,j)]]$ is irreducible, aperiodic and doubly stochastic.

\end{itemize}

\textbf{Remark:} The $i$th row $q(i, \cdot)$ of the  matrix $Q$ indicates the `weights' node $i$ assigns to its neighbors. Since it stands to reason that each node  values its own opinions, $q(i, i) > 0$, which automatically ensures aperiodicity.\\

Rewrite above iteration as
\begin{eqnarray}
r^{i}_{t+1} &=& r^{i}_t + {\gamma}_t [{\phi}^{i}(X_t) \overline{c}(X_t) + {\alpha}{\phi}^{i}(X_t)(\sum_{j=1}^{n} q(i,j)\sum_{s=1}^{m}p(X_t,s){\phi}^{j}(s)^{T}r^{j}_{t}) \nonumber \\
&&- \ {\phi}^{i}(X_t){\phi}^{i}(X_t)^{T}r^{i}_{t} + M^{i}_{t+1}], \label{iteration}
\end{eqnarray}
where $\overline{c}(i) = \sum_jp(i,j)c(i, j)$, $\overline{c} = [\overline{c}(1), \overline{c}(2), ....., \overline{c}(m)]^{T}$, and $M^{i}_{t+1}, t \geq 0,$ is a martingale difference sequence w.r.t.\ ${\sigma}(X_n,Y_n, n \le t)$, given by
\begin{eqnarray*}
M^i_{t + 1} &:=& \left(c(X_t,X_{t+1}){\phi}^{i}(X_t) - \overline{c}(X_t){\phi}^{i}(X_t)\right) +  {\alpha}\Big({\phi}^{i}(X_t){\phi}^{Y^{i}_{t+1}}(X_{t+1})^{T}r^{Y^{i}_{t+1}}_{t} \\
&& - \ {\phi}^{i}(X_t)(\sum_{j=1}^{n} q(i,j)\sum_{s=1}^{m}p(X_t,s){\phi}^{j}(s)^{T}r^{j}_{t})\Big).
\end{eqnarray*}

We have:
\begin{eqnarray*}
\sum_{l=1}^{m}d(l){\phi}^{i}(l) \overline{c}(l) &=&  {{\Phi}^{i}}^{T} {D} {\overline{c}}, \\
\sum_{l=1}^{m}d(l){\phi}^{i}(l){\phi}^{i}(l)^{T}\overline{r}^{i}_t &=& {{\Phi}^{i}}^{T} {D} {{\Phi}^{i}}{\overline{r}^{i}_{t}}, \\
\sum_{j=1}^{n}q(i,j)\sum_{l=1}^{m}d(l){\phi}^{i}(l)\sum_{s=1}^{m}p(l,s){\phi}^{j}(s)^{T}\overline{r}^{j}_{t} &=& {{\Phi}^{i}}^{T} D P \sum_{j=1}^{n} q(i,j){{\Phi}^{j}}{\overline{r}^{j}_{t}}.
\end{eqnarray*}
By Corollary 8, p.\ 74, Borkar (2008), the o.d.e.\ corresponding to (\ref{iter-i}) is
\begin{equation}
\dot{{\overline{r}}^{i}} ={{\Phi}^{i}}^{T} {D} {\overline{c}} + {\alpha} {{\Phi}^{i}}^{T} D P \sum_{j=1}^{n} q(i,j){{\Phi}^{j}}\overline{r}^{j} - {{\Phi}^{i}}^{T} {D} {{\Phi}^{i}}{\overline{r}^{i}}. \label{ode1}
\end{equation}
Let $\bar{r} = [ \overline{r}^{1},\overline{r}^{2},.......,\overline{r}^{n}]^{T}$, the concatenation of all $r^i$'s. This satisfies the o.d.e.
\begin{eqnarray*}
\lefteqn{\dot{\bar{r}} =} \\
&&\begin{bmatrix}
. \\
. \\
. \\
{{\Phi}^{i}}^{T} {D} {\overline{c}} + {\alpha} {{\Phi}^{i}}^{T} D P \sum_{j=1}^{n} q(i,j){{\Phi}^{j}}\overline{r}^{j} - {{\Phi}^{i}}^{T} {D} {{\Phi}^{i}}{\overline{r}^{i}} \\
. \\
. \\
. \\
\end{bmatrix}\\
&=& \left[\begin{array}{c} . \\ . \\ . \\ {{\Phi}^{i}}^{T} {D} {\overline{c}} \\ . \\ . \\ . \end{array} \right] + {\alpha}\left[\begin{array}{c} . \\ . \\ . \\  {{\Phi}^{i}}^{T} D P \sum_{j=1}^{n} q(i,j){{\Phi}^{j}}\overline{r}^{j}  \\ . \\ . \\ . \end{array} \right] - \left [\begin{array}{c}\\ . \\ . \\   {{\Phi}^{i}}^{T} {D} {{\Phi}^{i}}{\overline{r}^{i}} \\ . \\ . \\ . \end{array} \right] \\
&=&\begin{bmatrix} {{\Phi}^{1}}^{T} {D} & \cdots & 0  \\  \vdots & \ddots & \vdots  \\   0 & \cdots & {{\Phi}^{n}}^{T} {D}\end{bmatrix}
 \left[\begin{array}{c} {\overline{c}} \\ . \\ . \\{\overline{c}} \\ . \\ . \\ {\overline{c}} \end{array} \right] %+
%{\alpha}\begin{bmatrix} {{\Phi}^{1}}^{T} {D} & \cdots & 0  \\  \vdots & \ddots & \vdots  \\   0 & \cdots & {{\Phi}^{n}}^{T} {D}\end{bmatrix}
%\left[\begin{array}{c}  P\sum_{j=1}^{n} q(1,j){{\Phi}^{j}}\overline{r}^{j} \\ . \\ . \\ P\sum_{j=1}^{n} q(i,j){{\Phi}^{j}}\overline{r}^{j}  \\ . \\ . \\
%  P\sum_{j=1}^{n} q(n,j){{\Phi}^{j}}\overline{r}^{j} \end{array} \right] - \\
- \begin{bmatrix} {{\Phi}^{1}}^{T} {D} & \cdots & 0  \\  \vdots & \ddots & \vdots  \\   0 & \cdots & {{\Phi}^{n}}^{T} {D}\end{bmatrix}
\begin{bmatrix} {{\Phi}^{1}}  & \cdots & 0  \\  \vdots & \ddots & \vdots  \\   0 & \cdots & {{\Phi}^{n}} \end{bmatrix}
\left [\begin{array}{c} {\overline{r}^{1}} \\ . \\ . \\ {\overline{r}^{i}} \\ . \\ . \\ {\overline{r}^{n}} \end{array} \right] \\
&& + \ {\alpha}\begin{bmatrix} {{\Phi}^{1}}^{T} {D} & \cdots & 0  \\  \vdots & \ddots & \vdots  \\   0 & \cdots & {{\Phi}^{n}}^{T} {D}\end{bmatrix}
\left[\begin{array}{c}  P\sum_{j=1}^{n} q(1,j){{\Phi}^{j}}\overline{r}^{j} \\ . \\ . \\ P\sum_{j=1}^{n} q(i,j){{\Phi}^{j}}\overline{r}^{j}  \\ . \\ . \\
  P\sum_{j=1}^{n} q(n,j){{\Phi}^{j}}\overline{r}^{j} \end{array} \right]  \\
\end{eqnarray*}
\begin{eqnarray*}
&=& \begin{bmatrix} {{\Phi}^{1}}^{T} {D} & \cdots & 0  \\  \vdots & \ddots & \vdots  \\   0 & \cdots & {{\Phi}^{n}}^{T} {D}\end{bmatrix}
 \left[\begin{array}{c} {\overline{c}} \\ . \\ . \\{\overline{c}} \\ . \\ . \\ {\overline{c}} \end{array} \right] \\
 && - \ \ \begin{bmatrix} {{\Phi}^{1}}^{T} {D} & \cdots & 0  \\  \vdots & \ddots & \vdots  \\   0 & \cdots & {{\Phi}^{n}}^{T} {D}\end{bmatrix}
\begin{bmatrix} {{\Phi}^{1}}  & \cdots & 0  \\  \vdots & \ddots & \vdots  \\   0 & \cdots & {{\Phi}^{n}} \end{bmatrix}
\left [\begin{array}{c} {\overline{r}^{1}} \\ . \\ . \\ {\overline{r}^{i}} \\ . \\ . \\ {\overline{r}^{n}} \end{array} \right] \\
&& + \ {\alpha}\begin{bmatrix} {{\Phi}^{1}}^{T} {D} & \cdots & 0  \\  \vdots & \ddots & \vdots  \\   0 & \cdots & {{\Phi}^{n}}^{T} {D}\end{bmatrix}
\begin{bmatrix} q(1,1)P  & \cdots & q(1,n)P  \\  \vdots & \ddots & \vdots  \\   q(n,1)P & \cdots & q(n,n)P \end{bmatrix}
\left [\begin{array}{c} {{\Phi}^{1}}{\overline{r}^{1}} \\ . \\ . \\{{\Phi}^{i}} {\overline{r}^{i}} \\ . \\ . \\{{\Phi}^{n}} {\overline{r}^{n}} \end{array} \right].
\end{eqnarray*}

Thus we get the following equation:
\begin{eqnarray*}
\begin{split}
\dot{\bar{r}} &=& \begin{bmatrix} {{\Phi}^{1}}^{T}  & \cdots & 0  \\  \vdots & \ddots & \vdots  \\   0 & \cdots & {{\Phi}^{n}}^{T} \end{bmatrix}
\begin{bmatrix} D & \cdots & 0  \\  \vdots & \ddots & \vdots  \\   0 & \cdots & D \end{bmatrix}
 \left[\begin{array}{c} {\overline{c}} \\ . \\ . \\{\overline{c}} \\ . \\ . \\ {\overline{c}} \end{array} \right] \ + \ \ \ \ \ \ \ \ \ \ \ \ \ \ \ \ \ \ \ \ \ \ \ \ \ \ \ \ \ \ \ \ \\
&& {\alpha}\begin{bmatrix} {{\Phi}^{1}}^{T}  & \cdots & 0  \\  \vdots & \ddots & \vdots  \\   0 & \cdots & {{\Phi}^{n}}^{T} \end{bmatrix}
\begin{bmatrix} D & \cdots & 0  \\  \vdots & \ddots & \vdots  \\   0 & \cdots & D \end{bmatrix}
\begin{bmatrix} q(1,1)P  & \cdots & q(1,n)P  \\  \vdots & \ddots & \vdots  \\   q(n,1)P & \cdots & q(n,n)P \end{bmatrix} \\
&& \times \begin{bmatrix} {{\Phi}^{1}}  & \cdots & 0  \\  \vdots & \ddots & \vdots  \\   0 & \cdots & {{\Phi}^{n}} \end{bmatrix}
\left [\begin{array}{c} {\overline{r}^{1}} \\ . \\ . \\ {\overline{r}^{i}} \\ . \\ . \\ {\overline{r}^{n}} \end{array} \right]  \\
&& - \  \begin{bmatrix} {{\Phi}^{1}}^{T}  & \cdots & 0  \\  \vdots & \ddots & \vdots  \\   0 & \cdots & {{\Phi}^{n}}^{T} \end{bmatrix}
\begin{bmatrix} D & \cdots & 0  \\  \vdots & \ddots & \vdots  \\   0 & \cdots & D \end{bmatrix}
\begin{bmatrix} {{\Phi}^{1}}  & \cdots & 0  \\  \vdots & \ddots & \vdots  \\   0 & \cdots & {{\Phi}^{n}} \end{bmatrix}
\left [\begin{array}{c} {\overline{r}^{1}} \\ . \\ . \\ {\overline{r}^{i}} \\ . \\ . \\ {\overline{r}^{n}} \end{array} \right] \ \ \ \ \ \ \ \ \ \ \ \ \ \ \ \ \ \
\end{split}
\end{eqnarray*}

Consider an augmented state space  $S' := \{1,2,..n\} \times S$.  Order it as
\begin{displaymath}
\{(1,1),(1,2),...(1,m),(2,1),...(2,m),.....,(n,1),....,(n,m)\}.
\end{displaymath}
Define
\begin{eqnarray*}
\tilde{p}((i,x),(j,y)) &:=& q(i,j) \times p(x,y),  \\
\tilde{c}((i,x),(j,y)) &:=& c(x,y), \\
{\psi}_{jk}((i,x)) &:=& {\Phi}^{i}_{k}(x) \ \mbox{if} \ j = i, \ \mbox{else} \  0, \\
{\Psi} &:=& [{\psi}_{11}, {\psi}_{12},...{\psi}_{1{n_1}},......{\psi}_{n1},{\psi}_{n2},....{\psi}_{n{n_n}}]
\end{eqnarray*}
\begin{eqnarray*}
{\rho} &:=& \Big[\Big[\tilde{P}((i,x),(j,y))\Big]\Big] =\begin{bmatrix} q(1,1)P  & \cdots & q(1,n)P  \\  \vdots & \ddots & \vdots  \\   q(n,1)P & \cdots & q(n,n)P \end{bmatrix}, \\
{\Psi} &:=& \begin{bmatrix} {{\Phi}^{1}}  & \cdots & 0  \\  \vdots & \ddots & \vdots  \\   0 & \cdots & {{\Phi}^{n}} \end{bmatrix}, \\
{\Psi}^{T} &:=& \begin{bmatrix} {{\Phi}^{1}}^{T}  & \cdots & 0  \\  \vdots & \ddots & \vdots  \\   0 & \cdots & {{\Phi}^{n}}^{T} \end{bmatrix},
\end{eqnarray*}
\begin{eqnarray*}
{\nu} &:=& \frac{1}{n}\begin{bmatrix} D & \cdots & 0  \\  \vdots & \ddots & \vdots  \\   0 & \cdots & D \end{bmatrix}, \\
{\tilde{c}} &:=& \left [\begin{array}{c} \\ . \\ . \\ . \\ E[\tilde{c}((i,x),(j,y))|(i,x)] \\ . \\ . \\ . \end{array} \right] = \left [\begin{array}{c} \overline{c} \\ . \\ . \\ \overline{c} \\ . \\ . \\ \overline{c} \end{array} \right].\\
\end{eqnarray*}
Then the ODE for $r(\cdot)$ is
\begin{equation}
\dot{r} = n\left({\Psi}^{T}{\nu} \tilde{c} + {\alpha}{\Psi}^{T}{\nu}{\rho}{\Psi}r - {\Psi}^{T}{\nu}{\Psi}r\right). \label{ode}
\end{equation}

\textbf{Lemma 1:}
${\Psi}$ is a full rank matrix. \\

\textbf{Proof:} This is immediate from \textbf{(A1)}. \hfill $\Box$\\

\textbf{Lemma 2:}
$\rho$ is irreducible (hence positively recurrent) and aperiodic under \textbf{(A2)-(A3)}.\\

\textbf{Proof:} Let $p^{(n)}(i, j), q^{(n)}(k, \ell), \tilde{p}^{(n)}((k,i), (\ell, j))$ denote the $n$-step probabilities of going from $i$ to $j$, $k$ to $\ell$, $(k, i)$ to $(\ell, j)$ resp. for $n \geq 1$. Since $P, Q$ are irreducible aperiodic, there exist $n_0, n_0'$ such that $p^{(n)}(i, j) > 0, q^{(n')}(k, \ell) > 0$ for $n \geq n_0, n' \geq n_0'$ resp. So for $n \geq n_0\vee n_0'$, $\tilde{p}^{(n)}((k, i), (\ell, j)) > 0$. The claim follows. \hfill $\Box$ \\

Let $(Z_t, X_t), t \geq 0$, denote the augmented Markov chain with transition matrix $\rho$. Note that the diagonal entries of ${\nu}$ are $> 0$ and are the stationary probabilities under $\rho$, i.e., letting $\eta$ denote the  ordered vector thereof,  ${\eta}$ is a unique stationary distribution under $\nu$. \\

\textbf{Theorem 2} As $t\uparrow\infty$, $r_t, t \geq 0$, a.s.\ converges to an $r^*$ given as the unique solution to
\begin{displaymath}
{\Psi}^{T}{\nu} \tilde{c} + {\alpha}{\Psi}^{T}{\nu}{\rho}{\Psi}r^* - {\Psi}^{T}{\nu}{\Psi}r^* = 0.
\end{displaymath}
\ \\

\textbf{Proof:} The scalar $n$ on the right hand side of (\ref{ode}) does not affect its asymptotic behavior, so can be ignored. But then (\ref{ode}) is exactly of the same form as (\ref{ode0})  with the same assumptions being satisfied. Hence the same analysis applies. \hfill $\Box$\\

\textbf{Remark:} As in Tsitsiklis and Van Roy (1997), this can be extended to a positive recurrent Markov chain $\{X_t\}$ on a countably infinite state space under additional square-integrability assumptions on $\{c(X_t, X_{t+1}), \phi^i(X_t)\}$.\\

\section{Average cost}

Consider the problem of estimating average cost and a differential cost function on a finite, irreducible and aperiodic Markov chain. The average cost ${\mu}^{*}$ is given by $E_s[c(X_t,X_{t+1})]$, where $E_s[ \ \cdot \ ]$ denotes the stationary distribution. Let $\bar{\textbf{1}}$ denote a vector with all components equal to $1$. A differential cost function is any function $J: S \rightarrow R$ that satisfies the Poisson equation, which takes the form
\begin{displaymath}
J = \bar{c}-{\mu}^{*}\bar{\textbf{1}}+ PJ.
\end{displaymath}
It is known that for an irreducible Markov chain, differential cost functions exist and the set of all differential cost functions takes the form $\{J^{*} + c\bar{\textbf{1}}| c \in R \}$, for some $J^*$  satisfying $\eta^TJ^{*} = 0$. Such a $J^* $ is referred to as the basic differential cost function.  The original $TD(0)$ algorithm  for approximate evaluation of $J$ begins with the a priori approximation $J(\cdot) \approx \phi(\cdot) r = \sum_{i=1}^Kr_i\phi_i(\cdot)$. Here
 \begin{displaymath}
 {\phi}(X_t) = [{\phi}_{1}(X_t), {\phi}_{2}(X_t), .....{\phi}_{K}(X_t)]^{T},
  \end{displaymath}
 with ${\phi}_{i} :=$  the $ith$ feature vector (these are kept fixed), and $r := [r_1, \cdots, r_K]^T$ are the weights that are to be learnt. The actual algorithm for doing so is as follows Tsitsiklis and Van Roy (1999):
\begin{eqnarray*}
r_{t+1} &=& r_t + {\gamma}_t {\phi}(X_t) [c(X_t,X_{t+1}) - {\mu}_t + {\phi}(X_{t+1})^{T}r_{t} -{\phi}(X_t)^{T}r_{t}], \\
{\mu}_{t+1} &=& {\mu}_{t} + k{\gamma}_t(c(X_t,X_{t+1}) - {\mu}_t),
\end{eqnarray*}
where $k$ is any arbitrary positive constant. A convergence proof and error estimates relative to the exact $J$ may be found in Tsitsiklis and Van Roy (1999). As before, we sketch an alternative argument using the `o.d.e.' approach.
Once again, by Corollary 8, p.\ 74, Borkar (2008), the limiting o.d.e for the above iteration is
\begin{eqnarray*}
\dot{r} &=&  {\Phi}^{T}D\bar{c} - {\mu}{\Phi}^{T}\bar{\textbf{1}} + {\Phi}^{T}DP{\Phi}r -{\Phi}^{T}D{\Phi}r,  \\
\dot{\mu} &=& k({\mu}^{*} - {\mu}) \label{ode1}.
\end{eqnarray*}

Let $w_t = [{\mu}_t, r_t]^{T}$. In matrix notation, the o.d.e.\ can be written as
\begin{displaymath}
\dot{w} = \left[\begin{array}{c} \dot{\mu} \\ \dot{r} \end{array} \right ] = \begin{bmatrix} -k & {0  \cdots  0} \\ -{\Phi}^{T}D\bar{\textbf{1}} & {\Phi}^{T}DP{\Phi} -{\Phi}^{T}D{\Phi} \end{bmatrix}
\left[\begin{array}{c} {\mu} \\ r \end{array} \right ] +
\left[\begin{array}{c} k{\mu}^{*} \\ {\Phi}^{T}D\bar{c} \end{array} \right] =: h(w)
\end{displaymath}
for $h(w) := \begin{bmatrix} -k & {0  \cdots  0} \\ -{\Phi}^{T}D\bar{\textbf{1}} & {\Phi}^{T}DP{\Phi} -{\Phi}^{T}D{\Phi} \end{bmatrix}w +  \left[\begin{array}{c} k{\mu}^{*} \\ {\Phi}^{T}D\bar{c} \end{array} \right]$. Then
\begin{displaymath}
h_{\infty}(w) := \lim_{a\uparrow\infty}\frac{h(aw)}{a} = \begin{bmatrix} -k & {0  \cdots  0} \\ -{\Phi}^{T}D\bar{\textbf{1}} & {\Phi}^{T}DP{\Phi} -{\Phi}^{T}D{\Phi} \end{bmatrix}w.
\end{displaymath}
It is easy to see that $\frac{1}{a}h(aw) \rightarrow h_{\infty}(w)$ uniformly on $R^{K+1}$.\\

Let
\begin{eqnarray*}
A &:= & \begin{bmatrix} -k & {0  \cdots  0} \\ -{\Phi}^{T}D\bar{\textbf{1}} & {\Phi}^{T}DP{\Phi} -{\Phi}^{T}D{\Phi} \end{bmatrix} \text{and} \\
b &:=& \left[\begin{array}{c} k{\mu}^{*} \\ {\Phi}^{T}D\bar{c} \end{array} \right].
\end{eqnarray*}
Suppose we assume ${\Phi}$ has linearly independent columns and ${\Phi}r \ne e$ for any $r \in R^{K}$ \\

\textbf{Theorem 3:} Under the above assumptions, $w_t \ra \hat{w}$  a.s., where $\hat{w}$ is the unique  solution to $h(w) = 0$.\\

\textbf{Proof:} For sufficiently large $k$, the matrix $A$ is negative definite as seen from Lemma 7 of Tsitsiklis and Van Roy (1999) ( $k$ corresponds to their $l$). Hence the `scaled o.d.e.' $\dot{w}(t) = h_{\infty}(w(t))$ is a linear system with the origin as its globally asymptotically stable equilibrium, in fact $V(w(t)) = \|w(t)\|^2$ is a Liapunov function. By Theorem 9, p.\ 75, Borkar (2008), $\sup_t\|w_t\| < \infty$ a.s. In turn, (\ref{ode0}) has $\hat{w}$ as its globally asymptotically stable equilibrium, again $V(w(t)) = \|w(t) - \hat{w}\|^2$ is a Liapunov function. This can be seen as follows
\begin{eqnarray*}
\frac{dV(w(t))}{dt} &=& (w(t) - \hat{w})^{T}(Aw(t) + b) \\
 &=& (w(t) - \hat{w})^{T}(Aw(t) + b - A \hat{w} - b) \\
&=& (w(t) - \hat{w})^{T}A(w(t)-\hat{w}) \\
&\le&  0.
\end{eqnarray*}
 with equality iff $w(t)=\hat{w}$ .
The claim follows by Theorem 2, p.\ 15, Borkar (2008). \hfill $\Box$\\

Consider a similar setting as section 2. The $i$th agent thus runs the following $n_{i}$ dimensional iteration
\begin{eqnarray*}
r^{i}_{t+1} &=& r^{i}_t + {\gamma}_t {\phi}(X_t) [c(X_t,X_{t+1}) - {\mu}_{t}+ {\phi}^{Y^{i}_{t+1}}(X_{t+1})^{T}r^{Y^{i}_{t+1}} -{\phi}^{i}(X_t)^{T}r^{i}_{t}],   \\
{\mu}_{t+1} &=& {\mu}_{t} +  k{\gamma}_{t}[c(X_{t},X_{t+1}) - {\mu}_t],
\end{eqnarray*}
where $k$ is an arbitrary positive constant. We show convergence of the combined iterates $[r^{1}_{t}, r^{2}_{t},....,r^{n}_{t}, {\mu}_t]$. \\

Rewrite the above iteration as
\begin{eqnarray*}
r^{i}_{t+1} &=& r^{i}_t + {\gamma}_t  [{\phi}(X_t)\overline{c}(X_t) - {\phi}(X_t){\mu}_{t}+ {\phi}(X_t)\sum_{j=1}^{n}q(i,j) \sum_{s=1}^{m}p(X_t,s){\phi}^{j}(s)^{T}r^{j} \\
&& - \ {\phi}(X_t){\phi}^{i}(X_t)^{T}r^{i}_{t} + M^{i}_{t+1}],   \\
{\mu}_{t+1} &=& {\mu}_{t} +  k{\gamma}_{t}[\overline{c}(X_{t}) - {\mu}_t + M_{t+1}].
\end{eqnarray*}
Here $M^{i}_{t+1}$ and $M_{t+1}$ are martingale difference sequences given by resp.,
\begin{eqnarray*}
&& \left({\phi}^{i}(X_t){\phi}^{Y^{i}_{t+1}}(X_{t+1})^{T}r^{Y^{i}_{t+1}}_{t} - {\phi}^{i}(X_t)(\sum_{j=1}^{n} q(i,j)\sum_{s=1}^{m}p(X_t,s){\phi}^{j}(s)^{T}r^{j}_{t})\right) \\
&+&  c(X_t,X_{t+1}){\phi}^{i}(X_t) - \overline{c}(X_t){\phi}^{i}(X_t).
\end{eqnarray*}
and
\begin{displaymath}
k[c(X_t,X_{t+1}) - \overline{c}(X_t)].
\end{displaymath}

Using similar matrix notation from section 2 and using the fact that $\sum_{l=1}^{m}d(l)\overline{c}(l) = {\mu}^{*}$, the o.d.e.\ corresponding to (\ref{iteration}) is
\begin{eqnarray*}
\dot{\bar{r}}^{i} &=& {{\Phi}^{i}}^{T} {D} {\overline{c}} -\overline{\mu} {{\Phi}^{i}}^{T}D\bar{\textbf{1}}+ {{\Phi}^{i}}^{T} D P \sum_{j=1}^{n} q(i,j){{\Phi}^{j}}\bar{r}^{j} - {{\Phi}^{i}}^{T} {D} {{\Phi}^{i}}{\bar{r}^{i}},  \\
\dot{\overline{\mu}} &=& k({\mu}^{*} - \overline{\mu}).
\end{eqnarray*}

Let $r = [\bar{r}^{1},\bar{r}^{2},.....,\bar{r}^{n}]$, the concatentation of all $\bar{r}^{i}$'s. It satisfies the o.d.e.
\begin{eqnarray*}
\begin{split}
\dot{r} &=& \begin{bmatrix} {{\Phi}^{1}}^{T}  & \cdots & 0  \\  \vdots & \ddots & \vdots  \\   0 & \cdots & {{\Phi}^{n}}^{T} \end{bmatrix}
\begin{bmatrix} D & \cdots & 0  \\  \vdots & \ddots & \vdots  \\   0 & \cdots & D \end{bmatrix}
 \left[\begin{array}{c} {\overline{c}} \\ . \\ . \\{\overline{c}} \\ . \\ . \\ {\overline{c}} \end{array} \right] \ - \ \ \ \ \ \ \ \ \ \ \ \ \ \ \ \ \ \ \ \ \ \ \ \ \ \ \ \ \ \ \ \ \\
&& \overline{{\mu}}\begin{bmatrix} {{\Phi}^{1}}^{T}  & \cdots & 0  \\  \vdots & \ddots & \vdots  \\   0 & \cdots & {{\Phi}^{n}}^{T} \end{bmatrix}
\begin{bmatrix} D & \cdots & 0  \\  \vdots & \ddots & \vdots  \\   0 & \cdots & D \end{bmatrix}
 \left[\begin{array}{c} \bar{\textbf{1}} \\ . \\ . \\\bar{\textbf{1}} \\ . \\ . \\ \bar{\textbf{1}} \end{array} \right] \ + \ \ \ \ \ \ \ \ \ \ \ \ \ \ \ \ \ \ \ \ \ \ \ \ \ \ \ \ \ \ \ \ \\
\end{split}
 \end{eqnarray*}
 \begin{eqnarray*}
\begin{split}
&& \begin{bmatrix} {{\Phi}^{1}}^{T}  & \cdots & 0  \\  \vdots & \ddots & \vdots  \\   0 & \cdots & {{\Phi}^{n}}^{T} \end{bmatrix}
\begin{bmatrix} D & \cdots & 0  \\  \vdots & \ddots & \vdots  \\   0 & \cdots & D \end{bmatrix}
\begin{bmatrix} q(1,1)P  & \cdots & q(1,n)P  \\  \vdots & \ddots & \vdots  \\   q(n,1)P & \cdots & q(n,n)P \end{bmatrix} \\
&& \ \ \ \ \ \ \ \ \ \times \begin{bmatrix} {{\Phi}^{1}}  & \cdots & 0  \\  \vdots & \ddots & \vdots  \\   0 & \cdots & {{\Phi}^{n}} \end{bmatrix}
\left [\begin{array}{c} {\bar{r}^{1}} \\ . \\ . \\ {\bar{r}^{i}} \\ . \\ . \\ {\bar{r}^{n}} \end{array} \right].  \\
\end{split}
\end{eqnarray*}
\begin{eqnarray*}
\begin{split}
&& - \  \begin{bmatrix} {{\Phi}^{1}}^{T}  & \cdots & 0  \\  \vdots & \ddots & \vdots  \\   0 & \cdots & {{\Phi}^{n}}^{T} \end{bmatrix}
\begin{bmatrix} D & \cdots & 0  \\  \vdots & \ddots & \vdots  \\   0 & \cdots & D \end{bmatrix}
\begin{bmatrix} {{\Phi}^{1}}  & \cdots & 0  \\  \vdots & \ddots & \vdots  \\   0 & \cdots & {{\Phi}^{n}} \end{bmatrix}
\left [\begin{array}{c} {\bar{r}^{1}} \\ . \\ . \\ {\bar{r}^{i}} \\ . \\ . \\ {\bar{r}^{n}} \end{array} \right] .\ \ \ \ \ \ \ \ \ \ \ \ \ \ \ \ \ \
\end{split}
\end{eqnarray*}

Consider the augmented Markov chain as in section 2 and analogous definitions for ${\Psi}$,${\nu}$,${\rho}$ and $\tilde{c}$ .
Then the ODE for $r(\cdot)$ and $\overline{\mu}$ is ,
\begin{eqnarray}
\dot{r} &=& n\left({\Psi}^{T}{\nu} \tilde{c} - \overline{\mu}{\Psi}^{T}{\nu}e+ {\Psi}^{T}{\nu}{\rho}{\Psi}r - {\Psi}^{T}{\nu}{\Psi}r\right). \nonumber \\
\dot{\overline{\mu}} &=& k({\mu}^{*} - \overline{\mu}). \label{ode3}
\end{eqnarray}

We assume \textbf{A1, A2, A3 and A5} here as well. Hence Lemma 1, Lemma 2 hold in this case also. In addition we make the following key assumption.
\begin{itemize}
\item $\textbf{(A6)}$
${\Psi}r \ne e$ for any $r \in R^{n_1 + n_2 +...+n_n}$. \\
\end{itemize}

\textbf{Theorem 4} $\bar{\mu}_{t}$, $t \geq 0$ a.s.\ converges to ${\mu}^{*}$. $\bar{r}_t$, $t \geq 0$, a.s.\ converges to an $r^{*}$ , given as the unique solution to
\begin{displaymath}
{\Psi}^{T}{\nu} \tilde{c} -{\mu}^*{\Psi}^{T}{\nu}e+ {\Psi}^{T}{\nu}{\rho}{\Psi}r - {\Psi}^{T}{\nu}{\Psi}r = 0.
\end{displaymath}
\ \\

\textbf{Proof:} The scalar $n$ on the right hand side of (\ref{ode3}) does not affect its trajectory, so can be ignored. But then (\ref{ode3}) is exactly of the same form as (\ref{ode1}) with the same assumptions being satisfied. Hence the same analysis applies. \hfill $\Box$\\

\section{Discussion}

\begin{enumerate}

\item \textit{\large Performance comparison for Discounted Problem:}\\

Define $\Pi_i$ to be the projection onto the range of $\phi^i$ w.r.t.\ the weighted norm $\| \cdot \|$, where the weights are the values of the stationary probability distribtuion $\eta$. Let $T$ denote the Bellman operator defined by
\begin{displaymath}
(Tx)(i) := \bar{c}(i) + \alpha\sum_jp(i, j)x(j) \ \forall i.
\end{displaymath}
Recall from Tsitsiklis and Van Roy (1997) that this is a contraction w.r.t.\ the weighted norm above. Furthermore, by triangle inequality and convexity we have
\begin{eqnarray*}
\|J^*-\sum_{j=1}^{n}q(i,j)x_j\|^2 &\leq& \sum_{j=1}^{n}q(i,j)\|J^*-x_j\|^2. \\
\end{eqnarray*}
Let,
\begin{eqnarray*}
e^{*}_{i} &:=& \| J^{*} - \Pi_{i}J^{*} \| \\
e_{i} &:=& \| J^{*} - \phi^{i}{r^{*}}^{i} \| \\
e^{*} &:=& [e^{*}_{1},...,e^{*}_{i},...,e^{*}_{n}]^{T} \\
{e^{*}}^{(2)} &:=& [{e^{*}}^{2}_{1},...,{e^{*}}^{2}_{i},...,{e^{*}}^{2}_{n}]^{T} \\
e &:=& [e_{1},...,e_{i},...,e_{n}]^{T} \\
e^{(2)}&:=&[e^{2}_{1},...,e^{2}_{i},...,e^{2}_{n}]^{T} \\
\end{eqnarray*}
Our analysis borrows ideas from Tsitsiklis and Van Roy (1997,1999).
\begin{eqnarray*}
\| J^{*} - {\phi}^{i} {{r}^{*}}^{i}\|^2 &=& \|J^{*} - {\Pi}_{i}J^{*} \|^2 + \| {\Pi}_{i}J^{*} - {\phi}^{i}{{r}^{*}}^{i}\|^2 \\
&=& \| J^{*} - {\Pi}_{i}J^{*} \|^2 + \| {\Pi}_{i}TJ^{*} - {\Pi}_{i}T\large(\sum_{j=1}^{n}q(i,j){\phi}^{j}{{r}^{*}}^{j}\large)\|^2 \\
&\leq& \| J^{*} - {\Pi}_{i}J^{*} \|^2 + \| TJ^{*} - T\large(\sum_{j=1}^{n}q(i,j){\phi}^{j}{{r}^{*}}^{j}\large)\|^2 \\
&\leq& \| J^{*} - {\Pi}_{i}J^{*} \|^2 + \alpha^2\| J^{*} - \sum_{j=1}^{n}q(i,j){\phi}^{j}{{r}^{*}}^{j}\|^2 \\
&\leq& \| J^{*} - {\Pi}_{i}J^{*} \|^2 + \alpha^2\sum_{j=1}^{n}q(i,j)\| J^{*} - {\phi}^{j}{{r}^{*}}^{j}\|^2,
\end{eqnarray*}
The first equality follows from Pythagoras theorem. The first and second inequalities follows from non-expansivity of $\Pi_{i}$ and contraction property of $T$, respectively. Thus we have
\begin{eqnarray*}
e^2_{i} &\leq& {e^{*}}^2_{i} + \alpha^2\sum_{j}q(i,j)e^2_{j} \\
\Rightarrow e^{(2)} &\leq&  {e^{*}}^{(2)} + \alpha^{2} Qe^{(2)} \\
\Rightarrow e^{(2)} &\leq&  (I-\alpha^2 Q)^{-1}{e^{*}}^{(2)} \\
&=&  \large(\sum_{k=0}^{\infty}{\alpha}^{2k}Q^k\large ) {e^{*}}^{(2)}.
\end{eqnarray*}
This is justified because the last expression shows that $(I-\alpha^2 Q)^{-1}$ is a non-negative matrix. Let $\tilde{Q} := (1-\alpha^2)\sum_{k=0}^{\infty}{\alpha}^{2k}Q^{k}$, a doubly stochastic matrix. Thus we have
\begin{eqnarray*}
e^{(2)} &\leq& (1-\alpha^2)^{-1} \tilde{Q}{e^{*}}^{(2)} \\
\Rightarrow \max_{i}e^2_{i} &\leq& \frac {\beta(e^*)}{(1-\alpha^2)} \max_{i}{e_{i}^{*}}^2 \\
\Rightarrow \max_{i}e_{i} &\leq& \frac {\sqrt{\beta(e^*)}}{\sqrt{(1-\alpha^2)}} \max_{i}{e_{i}^{*}}, \\
\end{eqnarray*}
where $\beta(e^*),\sqrt{\beta(e^*)} \in (0,1)$. The second inequality follows if we assume that an agent with the maximum $e^{*}_{i}$ samples an agent with a lesser $e^{*}_{i}$ with non-zero probability. This assumption in turn follows from irreducibility and an assumption that at least one $e^{*}_{i}$ is different from the rest. Thus we get a multiplicative improvement over $\frac{\max_{i}{e_{i}^{*}}}{\sqrt{(1-\alpha^2)}}$, which would correspond to  the estimate from Tsitsiklis and Van Roy (1997).
 Let $\Pi := \frac1 n\sum_{i=1}^{n}\Pi_{i}$ and $\bar{J} = \frac{1}{n}\sum_{i=1}^{n}\phi^{i}{r^{*}}^{i}$. Then
\begin{eqnarray*}
\|J^{*}-\bar{J}\| &\leq& \|J^* - \Pi J^* \| + \|\Pi J^* - \bar{J} \| \\
&\leq& \|J^* - \Pi J^* \| + \frac 1 n \sum_{i=1}^{n}\|\Pi_{i} J^* - \phi^{i}{r^{*}}^{i} \| \\
&\leq& \|J^* - \Pi J^* \| + \frac 1 n \sum_{i=1}^{n}\alpha \sum_{j=1}^{n} q(i,j) \| J^* - \phi^{j}{r^{*}}^{j} \| \\
&\leq& \|J^* - \Pi J^* \| + \frac 1 n \sum_{i=1}^{n}\alpha  \| J^* - \phi^{i}{r^{*}}^{i} \| \\
&\leq& \|J^* - \Pi J^* \| + \frac {\alpha\beta(e^*)}{(1-\alpha)} \max_{i}e^*_{i} \\
&\leq& \frac{(1-\alpha)\|J^* - \Pi J^* \| + {\alpha\beta(e^*)} \max_i e^*_{i}}{(1-\alpha)}. \\
\end{eqnarray*}
The numerator is a convex combination of $\|J^* - \Pi J^* \|$ and $\beta(e^*) \max_{i}e^*_{i}$, a multiplicative improvement over $ \max_i e^*_{i}$.\\

This suggests that the maximum error and the variance should be less in the distributed algorithm. We do not, however, have a formal proof for this. We have instead included some simulations that support this intuition. \\

Specifically, we have considered the problem of calculating the average discounted sum of queue lengths over an infinite horizon. The maximum queue length is capped at $50$. The arrival probability is $0.3$ and the departure probability is $0.35$. The discount factor is $0.9$. We have considered $3$ agents. The sampling probabilities, basis functions and initial values of weights are as follows:

\begin{eqnarray*}
Q &=& \begin{bmatrix} 5/12& 5/12& 1/6 \\  1/4&1/4&1/2  \\ 1/3&1/3&1/3 \end{bmatrix} \\
\phi^{1}_{1}(i) &=& I\{i > 5 \} \\
\phi^{1}_{2}(i) &=& I\{i > 10 \} \\
\phi^{1}_{3}(i) &=& I\{i > 20 \} \\
\phi^{1}_{4}(i) &=& \frac {i}{\frac{1}{51}\sum_{k=0}^{50}k} \\
\phi^{2}_{1}(i) &=& I\{|i-25| < 5 \} \\
\phi^{2}_{2}(i) &=& I\{|i-35| < 10 \} \\
\phi^{2}_{3}(i) &=& \frac{i^2}{\frac{1}{51}\sum_{k=0}^{50}k^2} \\
\phi^{3}_{1}(i) &=& \frac{\sqrt{i}}{\frac{1}{51}{\sum_{k=0}^{50}\sqrt{k}}} \\
\phi^{3}_{2}(i) &=& I\{i > 30 \} \\
r^{i}_0 &=& [0,0,.....n_i   \text{ times}...., 0..0]^{T} \\
\end{eqnarray*}

The plot of variance vs iteration and maximum error vs iteration for both our distributed algorithm (coupled) and the uncoupled algorithm are given below. We see that the variance and maximum error is lower in the former case.  \\

\begin{figure}[H]
\centering
 \includegraphics[scale = 0.5]{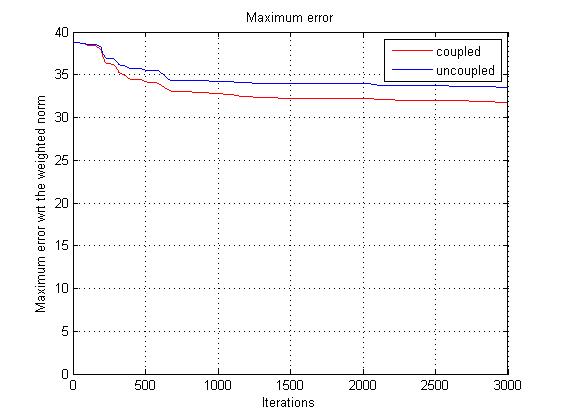} \\
\caption{Maximum error vs number of iterations}
\end{figure}
\begin{figure}[H]
\centering
 \includegraphics[scale = 0.5]{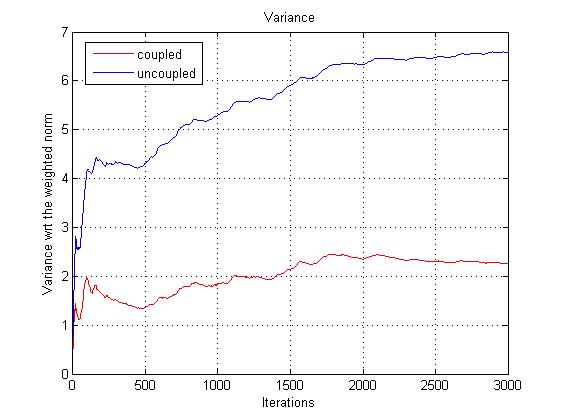} \\
\caption{Variance vs Number of iterations}
\end{figure}

\item \textit{\large Performance comparison for Average Cost Problem:}\\

Define the equivalence relation $i' \approx j'$ if for some finite $n \geq 1$, there is a sequence $i' = i_0, j_0, i_1, j_1, \cdots, i_{n-1}, j_{n-1}, i_n = j'$ such that $p(j_k, i_k)$, $p(j_k, i_{k+1}) > 0 \ \forall \ 0 \leq k < n$. Consider equivalence classes under $\approx$. We assume that the whole state space is one equivalence class. Note that this assumption will be satisfied if every node has a self loop, which is true for most queuing models. Furthermore, it does not cause any loss of generality as observed in Tsitsiklis and Van Roy (1999), because inserting a self-loop of probability $\delta \in (0,1)$ at each state is tantamount to replacing $P$ by $(1 - \delta)P + \delta I$, equivalently, introducing a sojourn time binomially distributed with parameter $\delta$ at each state. This does not affect either $\beta$ or $J^*$, and amounts to a harmless time scaling for the algorithm. \\

Let $\theta$ denote the zero vector.\\

\textbf{Lemma 3} Under the above assumption, $\sup_{x \neq \theta, \eta^Tx = 0}\frac{\|Px\|}{\|x\|} < 1$, where $\| \ \cdot \ \|$ is the weighted norm.\\

\noindent \textbf{Proof} Clearly $\sup_{x \neq \theta, \eta^Tx = 0}\frac{\|Px\|}{\|x\|} \leq 1$. If equality holds, we must have
\begin{displaymath}
\sum_jp(i,j)x_j^2 = (\sum_jp(i,j)x_j)^2 \ \ \ \forall i,
\end{displaymath}
which is possible only if for each $i$, $x_j$ is constant for $j \in \{k : p(i, k) > 0\}$. Thus $x_{i'} = x_{j'}$, if $i'$ and $j'$ belong to the same equivalence class (defined above). Since we have assumed that the entire state space is a single equivalence class, $x$ is a constant vector. This along with $\eta^Tx=0, x \ne \theta$ gives us a contradiction. Hence the claim follows.  \hfill $\Box$\\

Define $\Pi_{i}$ is before. As before Let $\bar{\textbf{1}}:=[1,1,..,1,...,1]^T$.  Define $\Pi_{\bar{\textbf{1}}^c}$ to be the projection on the subspace that is orthogonal to $\bar{\textbf{1}}$ with respect to the weighted norm $\| .\|$. Let $T$ be defined as
\begin{equation*}
(Tx)(i) := \bar{c}(i) - \mu^* + \sum_{j=1}^{n}p(i,j)x(j).
\end{equation*}
Since $J^*+c\bar{\textbf{1}} \ \forall c \in R$ are valid differential cost functions, we define the error of the $i^{th}$ agent as:
\begin{eqnarray*}
\inf_{c \in R}\|J^*+c\bar{\textbf{1}} - \phi^{i}{r^{*}}^{i}\| \\
= \|\Pi_{\bar{\textbf{1}}^c}(J^* - \phi^{i}{r^{*}}^{i})\| \\
\end{eqnarray*}
As $\eta^TJ^* = 0$, we have $\Pi_{\bar{\textbf{1}}^c}J^*=J^*$ and hence the error is equal to
\begin{equation*}
\|J^* - \Pi_{\bar{\textbf{1}}^c}\phi^{i}{r^{*}}^{i} \|.
\end{equation*}
Let $e$, $e^{(2)}$, $e^*$ and ${e^*}^{(2)}$ be defined analogously. We assume that $ \forall i \ \phi^{i}_{j} \  j \in {1,2,..n_i}$ are orthogonal to $\bar{\textbf{1}}$ w.r.t.\ the weighted norm. This gives us:
\begin{eqnarray}
\eta^TJ^* &=&0 ,\nonumber \\
\eta^T\phi^{j} &=&0.  \label{ortho} \\ \nonumber
\end{eqnarray}
This assumption is not restrictive since $\eta^TJ^* = 0$ and we are approximating $J^*$.  This leads to
\begin{eqnarray*}
\Pi_{i}\mu^*\bar{\textbf{1}}&=&0 \\
\Rightarrow \Pi_{i}\mu^*\bar{\textbf{1}}&=&\Pi_{i}(\eta^{T}J^*)\bar{\textbf{1}}. \\
\Pi_iTJ^* &=& \Pi_i(c-\mu^*\bar{\textbf{1}}+ PJ^*) \\
&=& \Pi_i(c-(\eta^TJ^*)\bar{\textbf{1}}+ PJ^*) \\
&=& \Pi_i(c + (P - \bar{\textbf{1}}\eta^T)J^*). \\
\text{Similarly,} \  \ \Pi_i \phi^i{r^*}^i &=& \Pi_i(c + (P - \bar{\textbf{1}}\eta^T)\phi^i{r^*}^i). \\
\end{eqnarray*}
Hence
\begin{eqnarray*}
\|J^* - \Pi_{\bar{\textbf{1}}^c}\phi^{i}{r^{*}}^{i} \|^2 &=& \|J^* - \phi^{i}{r^{*}}^{i} \|^2 \\
&=&  \|J^* - \Pi_iJ^* \|^2 + \|\Pi_iJ^* - \phi^{i}{r^{*}}^{i} \|^2 \\
&=&  \|J^* - \Pi_iJ^* \|^2 + \|\Pi_iTJ^* - \Pi_iT\sum_{j=1}^{n}q(i,j)\phi^j{r^*}^j \|^2 .\\
\end{eqnarray*}
For the sake of brevity, denote $(J^*-\sum_{j=1}^{n}q(i,j)\phi^j{r^*}^j)$ by $E_i$. By (\ref{ortho}) $\eta^TE_i=0$. Thus,
\begin{eqnarray*}
\|J^* - \Pi_{\bar{\textbf{1}}^c}\phi^{i}{r^{*}}^{i} \|^2&=&  \|J^* - \Pi_iJ^* \|^2 + \|\Pi_i (P - \bar{\textbf{1}}\eta^T)E_i \|^2 \\
&\leq&  \|J^* - \Pi_iJ^* \|^2 + \|(P - \bar{\textbf{1}}\eta^T)E_i \|^2 \\
&\leq& \| J^{*} - {\Pi}_{i}J^{*} \|^2 +\sum_{l=1}^{m}\eta(l)\large(\sum_{j=1}^{m}(p(l,j)-\eta(j))E_i(j)\large)^2 \\
&=& \| J^{*} - {\Pi}_{i}J^{*} \|^2 +\sum_{l=1}^{m}\eta(l)\large(\sum_{j=1}^{m}(p(l,j)(E_i(j)-\eta^TE_i)\large)^2 \\
&\leq& \| J^{*} - {\Pi}_{i}J^{*} \|^2 +\alpha^2\sum_{l=1}^{m}\eta(l)\large(E_i(l)-\eta^TE_i \large)^2 \\
&\leq& \| J^{*} - {\Pi}_{i}J^{*} \|^2 +\alpha^2\|E_i-\eta^TE_i \|^2 \\
&=& \| J^{*} - {\Pi}_{i}J^{*} \|^2 +\alpha^2\|J^*-\sum_{j=1}^{n}q(i,j)\phi^j{r^*}^j \|^2,
\end{eqnarray*}
where $\alpha:=\sup_{\eta^Tx=0, x\ne 0}\frac{\|Px\|}{\|x\|}$. From  Lemma 3 we know that $\alpha < 1$. Following the steps in the preceding section we get,
\begin{eqnarray*}
\max_{i}e_{i} &\leq& \sqrt{\frac{\beta(e^*)}{1-\alpha^2}}\max_ie^*_i \\
\|J^{*}-\bar{J}\| &\leq& \frac{(1-\alpha)\|J^* - \Pi J^* \| + {\alpha\beta(e^*)} \max_i e^*_{i}}{(1-\alpha)}, \\
\end{eqnarray*}
where $\bar{J}$ and $\Pi$ are defined as before and $\beta(e^*) \in [0,1)$. Thus we get multiplicative improvements in the bound over the uncoupled case. It is worth noting that the bound derived in Tsitsiklis and Van Roy (1999) does not seem to extend easily to the distributed set-up. As before, we do expect that the variance  should be less in the distributed algorithm with gossip as opposed to the uncoupled case. Again we do not have a formal proof, but we have included simulations to support our intuition. We simulate with the same parameters in section 4.1 except that the feature vectors are projected on $\bar{\textbf{1}}^c$. The simulations showed significant reduction in variance, however the maximum error was approximately same for both. We have included the graph for variance here.
\begin{figure}[H]
\centering
 \includegraphics[scale = 0.5]{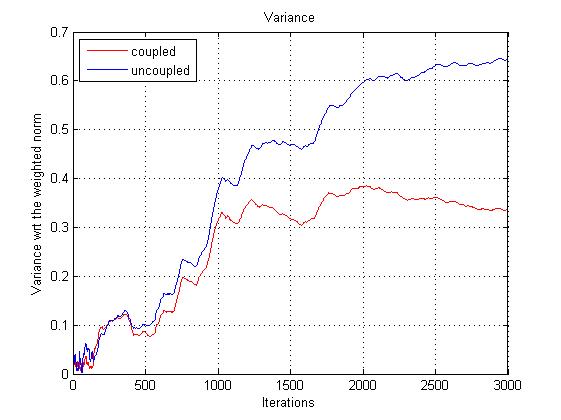} 	
\caption{Variance vs number of iteration}	
\end{figure}

\item Interestingly, the simple convergence proof above fails for $TD(\lambda)$ for $\lambda \neq 0$. It will be interesting to see whether our scheme can be modified to suit general $\lambda$.

\end{enumerate}


\begin{thebibliography}{99}

\bibitem{Bert} Bertsekas DP (2012) \textit{Dynamic Programming and Optimal Control, Vol.\ II} (4th edition), Athena Scientific, Belmont, MA

\bibitem{BorkarBook} Borkar VS (2008) \textit{Stochastic Approximation: A Dynamical Systems Viewpoint}, Hindustan Publ.\ Agency, New Delhi, India, and Cambridge Uni.\ Press, Cambridge, UK

\bibitem{Bus} Busoniu L, Babuska R,  De Schutter B (2008)``A comprehensive survey of multiagent reinforcement learning", \textit{IEEE Trans.\ on Systems, Man and Cybernetics, Part C: Applications and Reviews} 38, 156-172.

\bibitem{Der} Derevitskii DP, Fradkov AL (1974) ``Two models for analyzing the dynamics of adaptation algorithms", \textit{Automation and Remote Control} 35, 59-67.

\bibitem{Gosavi} Gosavi A (2003) \textit{Simulation-based Optimization, Parametric Optimization Techniques and Reinforcement Learning}, Springer Verlag, New York

\bibitem{Lauer} Lauer M, Riedmiller MA  (2000)``An algorithm for distributed reinforcement learning in cooperative multi-agent systems",  \textit{Proceeding of the Seventeenth International Conference on Machine Learning},
Morgan Kaufmann Publ., San Francisco, CA, 535-542.

\bibitem{Lewis} Lewis FL, Liu D\ (eds.) (2013) \textit{Reinforcement Learning and Approximate Dynamic Programming for Feedback Control}, Wiley, Hoboken, NJ

\bibitem{Litt} Littman M, Boyan J (1993) ``A distributed reinforcement learning scheme for network routing",  \textit{Proceedings of the 1993 International Workshop on Applications of Neural Networks to Telecommunications} (J.\ Alspector, R.\ Goodman, T.\ X.\ Brown, eds.), Lawrence Erlbaum Associates, Inc., Hillsdale, NJ, 45-51.

\bibitem{Llu} Llung L (1977) ``Analysis of recursive stochastic algorithms", \textit{IEEE Trans.\ on Automatic Control} 22, 551-575.

\bibitem{Macua} Macua SV, Belanovic P, Zazo S (2012) ``Diffusion gradient temporal difference for cooperative reinforcement learning with linear function approximation", \textit{Proceedings of the 3rd International Workshop on Cognitive Information Processing}, Parador de Baiona, Spain, 1-6.

\bibitem{Panait} Panait L, Luke S (2005) ``Cooperative multi-agent learning: the state of the art", \textit{Autonomous Agents and Multi-Agent Systems} 11, 387-434.

\bibitem{Pend} Pendrith MD (2000) ``Distributed reinforcement learning for a traffic engineering application", \textit{Proceedings of the Fourth International Conference on Autonomous Agents},
ACM, NY, 404-411.

\bibitem{Powell} Powell WH (2007) \textit{Approximate Dynamic Programming: Solving the Curses of Dimensionality}, Wiley, New York

\bibitem{Shah}  Shah D (2008) ``Gossip algorithms", \textit{Foundations and Trends in Networking}, Vol.\ 3(1), pp.\ 1-125

\bibitem{Szep} Szepesvari C (2010) \textit{Algorithms for Reinforcement Learning}, Morgan and Claypool Publishers

\bibitem{VanRoy} Tsitsiklis JN, Van Roy B (1997) ``An analysis of temporal-difference learning with function approximation", \textit{IEEE Trans.\ on Automatic Control} 42(5), 674-690.

\bibitem{VanRoy2} Tsitsiklis JN, Van Roy B (1999) ``Average cost temporal-difference learning", \textit{Automatica} 35, 1799-1808.

\bibitem{Weiss} Weiss G (1995) ``Distributed reinforcement learning", \textit{Robotics and Autonomous Systems} 15, 135-142.


\end{thebibliography}
\end{document}